\documentclass[aps,twocolumn,amssymb,amsmath,nofootinbib]{revtex4}

\usepackage{bbm}
\usepackage{graphicx}
\usepackage{subfigure}

\newcommand{\be}{\begin{eqnarray}}
\newcommand{\ee}{\end{eqnarray}}
\newcommand{\nn}{~\nonumber\\}
\newcommand{\sfr}[2]{{\textstyle\frac{#1}{#2}}}

\usepackage{color}

\newcommand{\intron}[1]{}

\renewcommand{\arraystretch}{1.5}

\begin{document}


\title{Walking in the SU(N)}

\author{Dennis D.~Dietrich}

\affiliation{Institute for Theoretical Physics, Heidelberg University,
Heidelberg, Germany}
\affiliation{The Niels Bohr Institute, Copenhagen, Denmark}

\author{Francesco Sannino}

\affiliation{Department of Physics and Chemistry,
University of Southern Denmark, Odense, Denmark}
\affiliation{The Niels Bohr Institute, Copenhagen, Denmark}


\begin{abstract}

We study the phase diagram as function of the number of colours and
flavours of asymptotically free non-supersymmetric theories with matter in
higher dimensional representations of arbitrary SU($N$) gauge groups. Since
matter in higher dimensional representations screens more than in the
fundamental a general feature is that a lower number of flavours is needed
to achieve a near-conformal theory. We study the spectrum of the theories
near the fixed point and consider possible applications of our analysis to
the dynamical breaking of the electroweak symmetry.

\end{abstract}


\maketitle


\section{Introduction}

In this article we portray the phase diagram, as a function of the number of
colours and the number of flavours, of non-supersymmetric gauge field theories 
with matter in higher dimensional representations of SU($N$) gauge groups. 
Strongly interacting non-Abelian theories generally feature a coupling 
constant which varies with the energy scale; it {\it runs}. This is caused by 
the antiscreening due to the charged gauge bosons (gluons). This 
antiscreening is competing with the screening contribution from matter fields. 
There is another feature of strongly interacting theories which comes into 
play in this context, that is chiral symmetry breaking. This formation of a 
$\langle\bar{\psi}\psi\rangle$ condensate below the chiral symmetry breaking 
scale renders the fermions massive and results in their decoupling from the
dynamics. In that case the antiscreening of the gluons becomes more
important again. With sufficient matter content an infrared fixed point of the
coupling constant can be reached, before chiral symmetry breaking is
triggered, that is there exists a conformal phase. For a number of flavours
slightly below the value for which the conformal phase is present, the
coupling constant is evolving slowly. It stays almost constant over a range
of energy scales. One says it {\it walks} instead of {\it runs}. These
features will be discussed in more detail below. 

We are particularly interested in these walking theories: It is possible
that, besides Quantum Chromodynamics (QCD), new strongly interacting
theories will emerge when exploring the unknown territory beyond the
Standard Model (SM) of particle interactions. For example, to avoid
unnaturally large quantum corrections to the mass scale of the electroweak
theory arising in the Higgs sector of the SM one can replace the elementary
Higgs by a strongly coupled sector. This approach has been named
technicolour \cite{TC}. The generation of the masses of the standard model fermions
requires extended technicolour interactions, which are consistent with
the observed amount of flavour-changing neutral currents and lepton number
violation for technicolour theories possessing a sufficient amount of 
walking
\cite{Holdom:1981rm,Yamawaki:1985zg,Appelquist:an,MY,Lane:1989ej}. 
The simplest of such models which also passes the electroweak
precision tests (like, for example the experimental bounds on the oblique
parameters) requires fermions in higher dimensional representations
of the technicolour gauge group
\cite{Sannino:2004qp,Hong:2004td,Dietrich:2005jn,Dietrich:2005wk}. Matter in
a higher dimensional representation screens more strongly than matter in the
fundamental representation, whence a smaller number of flavours is required
in order to achieve a given amount of screening. In the
recent past we have studied the gauge dynamics of fermions transforming
according to the two-index representations of SU($N$). Here we extend the
analysis to a generic asymptotically free gauge theory with fermions in
various higher dimensional representations of the SU($N$) group.
Some general structures are unveiled. 

After a general and fairly complete analysis (Sect. II) we explore the
various ways these theories can be used to break the electroweak symmetry
(Sect. III). Among other things we compute the left-right vector correlator 
in perturbation theory. Thanks to the results found in
\cite{Appelquist:1998xf,Appelquist:1999dq,Duan:2000dy} and corroborated
more recently by AdS/QCD-based computations \cite{Hong:2006si} we know that
near the conformal window the perturbative value of the oblique parameter $S$ 
provides a conservative estimate, that is an upper bound.
Finally, using the four-dimensional renormalisation group approach
we explore the spectrum near the fixed point (Sect. IV). In Sect.~V, we
summarise the results.


\section{Phase diagram}

\subsection{Review of the basic tools}

We start our journey by reviewing the two-loop $\beta$ function for a
generic non-Abelian gauge theory with fermionic matter in a given 
representation R of SU($N$)\cite{Caswell:1974gg}\footnote{Notice the
different normalisation as compared to, for example
\cite{Sannino:2004qp}. Conveniently, below, the quadratic Casimir operator
will only assume integer values. Nevertheless, many of the subsequent
expressions turn out to be identical due to the ratios that are taken.}:
\be
\beta(g)
&=&
-
\beta_0\sfr{g^3}{(4\pi)^2}
-
\beta_1\sfr{g^5}{(4\pi)^4},\\
2N\beta_0
&=&
\sfr{11}{3}C_2(\mathrm{G})-\sfr{4}{3}T(\mathrm{R}),
\label{beta0}
~\\
(2N)^2\beta_1
&=&
\sfr{34}{3}{C_2}^2(\mathrm{G})
-
\sfr{20}{3}C_2(\mathrm{G})T(\mathrm{R})-4C_2(\mathrm{R})T(\mathrm{R}).
\nn
\label{beta1}
\ee
$C_2(\mathrm{R})$ stands for the quadratic Casimir operator of the
representation R,
\be
2NX^a_\mathrm{R}X^a_\mathrm{R}=C_2(\mathrm{R})\mathbbm{1},
\ee
where $X^a_\mathrm{R}$ are the generators in the representation R. $C_2(\mathrm{G})$ is the quadratic Casimir
operator of the adjoint representation. $T(\mathrm{R})$ is the trace normalisation factor for the representation
R. $T(\mathrm{R})$ is connected to the quadratic Casimir operator, $C_2(\mathrm{R})$, by \cite{Jones:1981we}
\be
N_fC_2(\mathrm{R})d(\mathrm{R})=T(\mathrm{R})d(\mathrm{G}),
\label{tr}
\ee
where $d(\mathrm{R})$ denotes the dimension of the representation, R, and,
accordingly, $d(\mathrm{G})$ the dimension of the adjoint representation G
that is the number of generators. $N_f$ stands for the number of flavours.

A theory's loss of asymptotic freedom shows itself in a change of sign of the first coefficient, $\beta_0$, of
the $\beta$ function. Hence, the number of flavours, $N_f^\mathrm{I}[\mathrm{R}]$, above which the loss occurs
satisfies \mbox{$\beta_0[N_f^\mathrm{I}(\mathrm{R})]\overset{!}{=}0$}. Therefore, from Eqs.~(\ref{beta0}) and
(\ref{tr}) we obtain
\be
N_f^\mathrm{I}[\mathrm{R}]
:=
\frac{11}{4}\frac{d(\mathrm{G})C_2(\mathrm{G})}{d(\mathrm{R})C_2(\mathrm{R})},
\label{nfone}
\ee
which is thus proportional to the ratio of the second order indices,
$d(\mathrm{R})C_2(\mathrm{R})$, of the adjoint representation G
and the representation R, respectively. For matter in the adjoint
representation, $R=G$, the number of flavours above which asymptotic freedom
is lost is independent of the number of colours and equal to $\frac{11}{4}$.

Already for a smaller number of flavours than the one at which asymptotic
freedom is lost, the theory develops a Banks--Zaks infrared fixed point
\cite{Banks:1981nn}. It appears as soon as the second coefficient of the
$\beta$ function, $\beta_1$, changes sign, that is at
$\beta_1[N_f^\mathrm{III}(\mathrm{R})]\overset{!}{=}0$. Eqs.~(\ref{beta1}) and
(\ref{tr}) then lead to
\be
N_f^\mathrm{III}[\mathrm{R}]
=
\frac{d(\mathrm{G})C_2(\mathrm{G})}{d(\mathrm{R})C_2(\mathrm{R})}
\frac{17C_2(\mathrm{G})}{10C_2(\mathrm{G})+6C_2(\mathrm{R})}.
\label{bzfp}
\ee
This fixed point will not come into play if, before it is reached, the
coupling constant $\alpha$ becomes so large that chiral symmetry breaking
is triggered.

In this context, in order to determine the number of flavours
$N_f^\mathrm{II}[\mathrm{R}]$ above which the theory becomes conformal, we
employ the criterion derived in \cite{Appelquist:1988yc,Cohen:1988sq} and
follow the discussion in \cite{Sannino:2004qp}: Whether chiral symmetry can
be broken depends on the relative order (with respect to the energy scale)
of the values of the coupling constant at which a conformal fixed point is
encountered, $\alpha_*$, and where chiral symmetry breaks, $\alpha_c$. When
following the renormalisation group flow from higher to lower energies the
coupling $\alpha$ keeps growing as long as the $\beta$ function remains
finite and negative. If the value $\alpha_c$ is reached, the fermions
decouple, their screening effect is lost, and only the antiscreening of the
gauge bosons remains. In that case the conformal fixed point cannot be
reached. If, on the other hand, $\alpha_*$ is met before chiral symmetry is
broken, the coupling freezes (because the $\beta$ function becomes zero).
Now, in turn, the value $\alpha_c$ of the coupling required for chiral
symmetry breaking cannot be attained.

In \cite{Appelquist:1988yc,Cohen:1988sq}, the critical value $\alpha_c$ of
the coupling constant for which chiral symmetry breaking occurs is defined
as the value for which the anomalous dimension of the quark mass operator
becomes unity, $\gamma\overset{!}{=}1$. According to
\cite{Appelquist:1988yc}, in ladder approximation the critical value of the
coupling is given by \footnote{We investigate here whether chiral symmetry
breaking occurs in a {\it walking} theory. Thus, corrections to the ladder
approximation like vertex corrections, which take into account the change of
the coupling constant, are small. Alternative methods for calculating the
conformal window have been provided, e.g. in
\cite{Banks:1981nn,MY,Appelquist:1996dq,Braun:2006jd}.}
\be
\alpha_c=\frac{2\pi N}{3C_2(\mathrm{R})}.
\ee
Compared to that, the two-loop fixed point value of the coupling constant
reads \cite{Sannino:2004qp}
\be
\frac{\alpha^*}{4\pi}=-\frac{\beta_0}{\beta_1}.
\ee
For a fixed number of colours the critical number of flavours for which the
order of $\alpha_*$ and $\alpha_c$ changes is defined by imposing 
$\alpha_*\overset{!}{=}\alpha_c$, and is given by
\be
N_f^\mathrm{II}[\mathrm{R}]
&=&
\frac{d(\mathrm{G})C_2(\mathrm{G})}{d(\mathrm{R})C_2(\mathrm{R})}
\frac{17C_2(\mathrm{G})+66C_2(\mathrm{R})}{10C_2(\mathrm{G})+30C_2(\mathrm{R})}.
\label{ncrit}
\ee
In order to evaluate the above expressions and throughout the article we use
the Dynkin indices (the Dynkin labels of the highest weight of an
irreducible representation) to uniquely characterise the representations and
determine the relevant coefficients. We summarise, for the reader's
convenience, the relevant formulae in the appendix \ref{eleven}.


\subsection{Classification of asymptotically free theories}

As we increase the dimension of the representation of the matter fields the
screening effect of the matter compensates more and more the anti-screening
effect of the gauge bosons. Eventually one loses asymptotic freedom. Here we
will render quantitative this fact by constructing all of the theories which
are asymptotically free with at least two Dirac flavours.

For two flavours the fundamental, adjoint, as well as two-index symmetric and 
antisymmetric representations remain asymptotically free for any number of 
colours. We will show below that there are no exceptions from this rule for 
more than nine colours. Apart from theories based on these representations 
the only remaining variants are the three-index antisymmetric representation 
which is asymptotically free with two flavours for six to nine 
colours\footnote{For a number of colours less than six the three-index 
antisymmetric always coincides with another representation with a smaller 
number of indices.} and the four-index antisymmetric with two flavours for 
eight colours. For $N\le 9$ all such theories are listed in Table \ref{nine}.

\begin{table}[ht]
\begin{tabular}{c|c|c|c|c|c|c}
R&$\bar{\mathrm{R}}$&$N_f^\mathrm{I}$&$N_f^\mathrm{II}$&$N_f^\mathrm{III}$
&$\pi S$&$\lambda_*$\\
\hline
(1)&$\equiv$&11&$7\frac{73}{85}$&$5\frac{27}{49}$&1&3.229\\
(2)&$\equiv$&$2\frac{3}{4}$&$2\frac{3}{40}$&$1\frac{1}{16}$&$\frac{1}{2}$
&1,124.\\
\hline (10)&(01)&$16\frac{1}{2}$&$11\frac{32}{35}$&$8\frac{1}{19}$&$2\frac{1}{2}$
&10.33\\
(20)&(02)&$3\frac{3}{10}$&$2\frac{163}{325}$&$1\frac{28}{125}$&1&11.29\\
(11)&$\equiv$&$2\frac{3}{4}$&$2\frac{3}{40}$&$1\frac{1}{16}$&$1\frac{1}{3}$
&1,124.\\
\hline
(100)&(001)&22&$15\frac{361}{385}$&$10\frac{126}{205}$&$4\frac{2}{3}$&20.70\\
(200)&(002)&$3\frac{2}{3}$&$2\frac{82}{105}$&$1\frac{71}{201}$&$1\frac{2}{3}$
&5.865\\
(010)&$\equiv$&11&$8\frac{12}{115}$&$4\frac{52}{55}$&4&127,359.\\
(101)&$\equiv$&$2\frac{3}{4}$&$2\frac{3}{40}$&$1\frac{1}{16}$&$2\frac{1}{2}$
&1,124.\\
\hline (1000)&(0001)&$27\frac{1}{2}$&$19\frac{58}{61}$&$13\frac{32}{161}$
&$7\frac{1}{2}$&35.90\\
(2000)&(0002)&$3\frac{13}{14}$&$2\frac{747}{763}$&$1\frac{662}{1463}$
&$2\frac{1}{2}$&4.284\\
(0100)&(0010)&$9\frac{1}{6}$&$6\frac{191}{237}$&$3\frac{514}{537}$&5
&29.00\\
(1001)&$\equiv$&$2\frac{3}{4}$&$2\frac{3}{40}$&$1\frac{1}{16}$&4&1,124.\\
\hline
(10000)&(00001)&33&$23\frac{283}{295}$&$15\frac{123}{155}$&11&57.57\\
(20000)&(00002)&$4\frac{1}{8}$&$3\frac{33}{260}$&$1\frac{53}{100}$
&$3\frac{1}{2}$&2.519\\
(01000)&(00010)&$8\frac{1}{4}$&$6\frac{3}{20}$&$3\frac{21}{44}$&$7\frac{1}{2}$
&5,146.\\
(00100)&$\equiv$&$5\frac{1}{2}$&$4\frac{18}{145}$&$2\frac{14}{61}$ &$6\frac{2}{3}$&2,198.\\
(10001)&$\equiv$&$2\frac{3}{4}$&$2\frac{3}{40}$&$1\frac{1}{16}$&$5\frac{5}{6}$
&1,124.\\
\hline (100000)&(000001)&$38\frac{1}{2}$&$27\frac{584}{605}$&$18\frac{125}{317}$
&$30\frac{1}{3}$&87.67\\
(200000)&(000002)&$\frac{77}{18}$&$3\frac{2294}{9495}$&$1\frac{2168}{3663}$
&$4\frac{2}{3}$&3.061\\
(010000)&(000010)&$7\frac{7}{10}$&$5\frac{3186}{4225}$&$3\frac{356}{1825}$
&7&4.050\\
(001000)&(000100)&$3\frac{17}{20}$&$2\frac{8707}{9650}$&$1\frac{1941}{3890} $&$5\frac{5}{6}$
&4.421\\
(100001)&$\equiv$&$2\frac{3}{4}$&$2\frac{3}{40}$&$1\frac{1}{16}$&8&1,124.\\
\hline (1000000)&(0000001)&44&$31\frac{1537}{1585}$&$20\frac{828}{829}$&40
&128.6\\
(2000000)&(0000002)&$4\frac{2}{5}$&$3\frac{1141}{3425}$&$1\frac{851}{1325}$
&6&2.752\\
(0100000)&(0000010)&$7\frac{1}{3}$&$5\frac{829}{1695}$&$3\frac{7}{723}$
&$9\frac{1}{3}$&5.329\\
(0010000)&(0000100)&$2\frac{14}{15}$&$2\frac{8738}{39975}
$&$1\frac{1733}{15675}$&$9\frac{1}{3}$
&52.90\\
(1000001)&$\equiv$&$2\frac{3}{4}$&$2\frac{3}{40}$&$1\frac{1}{16}$
&$10\frac{1}{2}$&1,124.\\
\hline (10000000)&(00000001)&$49\frac{1}{2}$&$35\frac{326}{335}$&$23\frac{106}{157}$
&51&183.4\\
(20000000)&(00000002)&$4\frac{1}{2}$&$3\frac{516}{1265}$&$1\frac{1678}{2453}$
&$7\frac{1}{2}$&2.526\\
(01000000)&(00000010)&$7\frac{1}{14}$&$5\frac{1016}{3395}$
&$2\frac{1261}{1435}$&12&6.669\\
(10000001)&(10000001)&$2\frac{3}{4}$&$2\frac{3}{40}$&$1\frac{1}{16}$
&$13\frac{1}{3}$&1,124.\\
\end{tabular}
\caption{Complete list of asymptotically free SU($N$) theories with at least 
one family of fermions and up to nine colours. $\lambda_*$
[Eq.~(\ref{walking})] and $\pi S$ [Eq.~(\ref{S})] are calculated for 
$N_f<N_f^{II}$ and even.\label{nine}}
\end{table}

\renewcommand{\arraystretch}{2}
\begin{table*}[ht]
\begin{tabular}{c|c|c|c|c|c}
&$d(\mathrm{R})$&$C_2(\mathrm{R})$&$N^\mathrm{I}_f$&$N^\mathrm{II}_f$
&$N^\mathrm{III}_f$\\
\hline
{\bf F}&\tiny$N$&\tiny$N^2-1$&$\frac{11}{2}N$
&$\frac{2}{5}N\frac{50N^2-33}{5N^2-3}$
&$\frac{34N^3}{13N^2-3}$\\
{\bf G}&\tiny$N^2-1$&\tiny$2N^2$&$2\frac{3}{4}$&$2\frac{3}{40}$&$1\frac{1}{16}$\\
S$_n$&$\frac{(N+n-1)!}{n!(N-1)!}$
&\tiny$n(N-1)(N+n)$&$\frac{11N(n-1)!(N+1)!}{2(n+N)!}$
&$\frac{2N(33n^2(N-1)+33nN(N-1)+17N^2)n!(N+1)!}
       {5n[3n^2(N-1)+3nN(N-1)+2N^2](n+N)!}$
&$\frac{34N^3(n-1)!(N+1)!(n+N)!^{-1}}
       {3n^2(N-1)+3nN(N-1)+10N^2}$\\
{\bf S}$_2$&$\frac{N(N+1)}{2}$&\tiny$2(N-1)(N+2)$
&$\frac{11}{2}\frac{N}{N+2}$
&$\frac{N}{N+2}\frac{83N^2+66N-132}{20N^2+15N-30}$
&$\frac{17N^3}{(N+2)(8N^2+3N-6)}$\\
S$_3$&$\frac{N(N+1)(N+2)}{6}$&\tiny$3(N-1)(N+3)$&$\frac{11N}{(N+2)(N+3)}$
&$\frac{4N(-297+2N(99+58N))(N+1)!}{5(-27+N(18+11N))(N+3)!}$
&$\frac{68N^3(N+1)!}{(-27+N(18+19N))(N+3)!}$\\
A$_n$&$\frac{N!}{n!(N-n)!}$&\tiny$n(N-n)(N+1)$
&$\frac{11}{2}{\scriptstyle N}\binom{N-2}{n-1}^{-1}$
&$\binom{N-2}{n-1}^{-1}
\frac{(34+66n)N^3-66n(n-1)N^2-66n^2N}{(10+15n)N^2-15n(n-1)N-15n^2}$
&$\binom{N-2}{n-1}^{-1} \frac{34N^3}{10N^2 - 3n^2(N+1)+3nN(N+1)}$\\
{\bf A}$_2$&$\frac{N(N-1)}{2}$&\tiny$2(N-2)(N+1)$&$\frac{11}{2}\frac{N}{N-2}$
&$\frac{N}{N-2}\frac{83N^2-66N-132}{20N^2-15N-30}$
&$\frac{17N^3}{(N-2)(8N^2-3N-6)}$\\
A$_3$&$\frac{N(N-1)(N-2)}{6}$&\tiny$3(N-3)(N+1)$&$\frac{11N}{(N-2)(N-3)}$
&$\frac{4N}{(N-2)(N-3)}\frac{116N^2-198N-297}{55N^2-90N-105}$
&$n=3$ in A$_n$ above\\
A$_4$&$\frac{N(N-1)(N-2)(N-3)}{24}$&\tiny$4(N-4)(N+1)$
&$\frac{33N}{(N-2)(N-3)(N-4)}$
&$\frac{6N}{(N-2)(N-3)(N-4)}\frac{149N^2-396N-528}{35N^2-90N-120}$
&$n=4$ in A$_n$ above\\
R$_1$&$\frac{(N-1)N^2(N+1)}{12}$&\tiny$4(N^2-4)$&$\frac{33}{2(N^2-4)}$
&$\frac{3(149N^2 - 528)}{5(7N^4 - 52N^2 +96)} $
&$\frac{51N^2}{11N^4-68N^2 + 96}$\\
R$_2$&$\frac{(N-1)N(N+1)}{3}$&\tiny$3(N^2-3)$&$\frac{11N}{2(N^2-3)}$
&$\frac{2N(-297+116N^2)}{5(81-60N^2+11N^4)}$&$\frac{34N^3}{81-84N^2+19N^4}$
\\
R$_3$&$\frac{(N+1)N(N-2)}{2}$&\tiny$(3N+1)(N-1)$
&$\frac{11N}{(3N+1)(N-2)}$&
$\frac{4N(-33-66N+116N^2)}{5(6+27N-N^2-73N^3+33N^4)}$&
$\frac{68N^3}{6+27N-17N^2-113N^3+57N^4} $\\
R$_4$&$\frac{(N+1)N^2(N-3)}{4}$&\tiny$4N(N-1)$&$\frac{11}{2N(N-3)}$&
$\frac{149N-132}{5N(N-3)(7N-6)}$&
$\frac{17}{18-39N+11N^2}$
\end{tabular}
\caption{
Characteristic quantities sorted after representations:
F=fundamental, G=adjoint, S$_n$=$n$-index symmetric, A$_n$=$n$-index
antisymmetric, R$_1$=(020\dots0), R$_2$=(110\dots0), R$_3$=(10\dots010),
R$_4$=(010\dots010). Representations marked in boldface lead to theories
that stay asymptotically free with at least two flavours for any number of
colours.
}
\label{table:reps}
\end{table*}
\renewcommand{\arraystretch}{1.5}

That at a large number of colours only the fundamental, adjoint, two-index
symmetric, and two-index antisymmetric representations survive is due to the
fact that they are the only ones whose dimension grows quadratically or
more slowly with the number of colours. It can thus be compensated by the
quadratic growth of the dimension of the adjoint representation in the
expressions for $N_f^\mathrm{I}$. For any given representation the quadratic
Casimir operators grow quadratically with (large) $N$ and their ratio for
different representations goes to a constant.


\subsubsection{Only four remaining representations for $N>9$}

In this subsection we prove that\\

{\it There exist no asymptotically free
theories with at least two flavours and ten or more colours which are not
contained in the fundamental, adjoint, or the two-index
representations.}\\

\noindent
The utilised methods are general and can be applied to different numbers of
Dirac flavours. We will consider also the case of one flavour. We start by
analysing the ordering of $N_f^\mathrm{I}(\mathrm{R})$ at fixed $N$ for
different representations, R.


\subsubsection*{Ordering $N_f^\mathrm{I}(\mathrm{R})$ at fixed $N$}

As all coefficients in Eqs.~(\ref{C}) and (\ref{d}) are
positive, we have, at a fixed number of colours, $N$ (that is with a fixed
number of Dynkin indices),
\be
a_j\ge b_j~\forall~j\in\{1;\dots;N\}
&\Rightarrow&
C_2(\{a_j\})\ge C_2(\{b_j\})
\nn
&\mathrm{and}&
d(\{a_j\})\ge d(\{b_j\}).
\nn
\ee
Using Eq.~(\ref{nfone}) we find
\be
a_j\ge b_j~\forall~j\in\{1;\dots;N\}
&\Rightarrow&
N_f^\mathrm{I}(\{a_j\})\le N_f^\mathrm{I}(\{b_j\}).
\nn
\ee
By increasing the value of any of the Dynkin indices the critical number of
flavours decreases. The next step is to determine a complete set of theories 
with $N_f^\mathrm{I}(R)$ just below two. They are to serve as universal
bounds on $N_f^\mathrm{I}(R)$ for all theories.


\subsubsection*{$n\ge 3$-index antisymmetric representations}

Let us start from the $n\ge 3$-index antisymmetric representations,
$\mathrm{A}_n(N)$.  We notice from the corresponding expression for
$N_f^\mathrm{I}[\mathrm{A}_n(N)]$ in
Table \ref{table:reps} that for $N\ge 2n$, \footnote{We are in the
right-hand flank of Pascal's triangle. For $N<2n$ the $n$-index
antisymmetric representation is covered by an $(n^\prime<n)$-index
antisymmetric representation due to the symmetry of the result under
conjugation, that is inversion of the order of the Dynkin indices. For
example (0010), which is A$_3$, is, in this sense, equivalent to
$\overline{(0010)}=(0100)$ which is A$_2$.}
\be
N_f^\mathrm{I}[\mathrm{A}_n(N)]
>
N_f^\mathrm{I}[\mathrm{A}_n(N+1)]
\ee
and
\be
N_f^\mathrm{I}[\mathrm{A}_{n-1}(N)]
>
N_f^\mathrm{I}[\mathrm{A}_n(N)].
\ee
We deduce that $N_f^\mathrm{I}$ shrinks if the number of colours, $N$, or
the number of indices is increased as long as \mbox{$N\ge 2n$}.

Therefore, given a reference value $N_f^\mathrm{I}[\mathrm{A}_{n_0}(N_0)]$,
\be
(n>n_0~\mathrm{or}~N>N_0)
\Rightarrow
N_f^\mathrm{I}[\mathrm{A}_{n}(N)]
<
N_f^\mathrm{I}[\mathrm{A}_{n_0}({N_0})].
\nn
\ee

Picking as reference value $N_f^\mathrm{I}[\mathrm{A}_3({10})]=1\frac{27}{28}$,
we know that
\be
N_f^\mathrm{I}[\mathrm{A}_n({N})]\le 1\sfr{27}{28}~~\forall~~n\ge 3,~N\ge 10.
\ee
For a given number of colours, $N$, starting from all antisymmetric
representations from A$_3(N)$ to A$_{N-3}(N)$, we can reach other
representations by {\it increasing} the values of the Dynkin indices.
The Dynkin representation of an $n$-index antisymmetric representation reads
\mbox{$(0\dots0\underset{n}{1}0\dots0)$}.
Therefore, the only representations that cannot be reached in this way are
\mbox{$(a_1a_20\dots0a_{N-2}a_{N-1})$} $\forall~a_1,a_2,a_{N-2},a_{N-1}$.
In combination with the findings of the previous subsection this leads to the
conclusion that only these latter representations may, but need not, have
$N_f^\mathrm{I}[\mathrm{A}_n(N)]>1\sfr{27}{28}$.


\subsubsection*{S$_3$ and (020\dots000)}

As can be seen from Table \ref{nine} we cannot obtain an equally low boundary
from the fundamental, (100\dots000), and the two-index antisymmetric
representation, (010\dots000). The same holds for the two-index symmetric
representation, (200\dots000). Hence, we continue with the three-index
symmetric representation, S$_3(N)$=(300\dots000), and the representation
R$_1(N)$:=(020\dots000). $N_f^\mathrm{I}[\mathrm{S}_3(N)]$ and
$N_f^\mathrm{I}[\mathrm{R}_1(N)]$ are monotonically decreasing functions of the
number of colours. This can be seen from the explicit expressions in
Table \ref{table:reps}.
As reference values, we can pick
$N_f^\mathrm{I}[\mathrm{S}_3(3)]=1\sfr{1}{10}$
and $N_f^\mathrm{I}[R_1(4)]=1\sfr{3}{8}$, which are both smaller than the
previous reference value
$N_f^\mathrm{I}[\mathrm{A}_3({10})]=1\frac{27}{28}$ and are situated at a
smaller number of colours as well.
The representations of the set
\mbox{$\{(a_1a_20\dots0a_{N-2}a_{N-1})$}
$\forall~a_1,a_2,a_{N-2},a_{N-1}\in\mathbbm{N}\}$,
which cannot be obtained by increasing the Dynkin indices either of S$_3(N)$
or of R$_1(N)$ or of their conjugates are given by
\mbox{$(a_1a_20\dots0a_{N-2}a_{N-1})$}
$\forall$ (\mbox{$a_1,a_{N-1}\in\{0;1;2\}$} and
\mbox{$a_2,a_{N-2}\in\{0;1\}$}). This amounts to 36 combinations before
making use of symmetry properties under conjugation.


\subsubsection*{(110\dots000), (100\dots010), and (010\dots010)}

In the next step, we exploit information on the representations
R$_2(N)$=(110\dots000), R$_3(N)$=(100\dots010), and R$_4(N)$=(010\dots010) in order
to set limits for
most of the remaining representations. To this end, we repeat the same steps
as before. First we check that $N_f^\mathrm{I}$ for these representations is
a decreasing function of the number of colours. This can be seen directly
from the corresponding expressions in Table \ref{table:reps}.
Subsequently, we pick reference values for the three cases:
$N_f^\mathrm{I}[\mathrm{R}_2(4)]=1\frac{9}{13}$,
$N_f^\mathrm{I}[\mathrm{R}_3(4)]=1\frac{9}{13}$,
and
$N_f^\mathrm{I}[\mathrm{R}_4(4)]=1\frac{3}{8}$, respectively. They are all
smaller than $N_f^\mathrm{I}[\mathrm{A}_3({10})]=1\frac{27}{28}$ and lie at a
smaller number of colours. The limits set by these reference values cannot be
exceeded for a larger number of colours in these representations. Finally,
we eliminate all representations from the remaining 36 that can be
generated by adding to the Dykin indices of R$_2$, R$_3$, or R$_4$. This
leaves but F, G, A$_2$, and S$_2$ as well as their conjugates
$\overline{\mathrm{F}}$, $\overline{\mathrm{A}_2}$, and
$\overline{\mathrm{S}_2}$ (G is a real representation), that is seven
representations of which four are independent.

Thereby we have shown that beyond ten colours no other representations lead
to asymptotically free theories with two flavours but the fundamental, the
adjoint, as well as the two-index symmetric and antisymmetric
representations. Below ten colours the only exceptions are given by the
three-index antisymmetric representation at six to nine colours and the
four-index antisymmetric at eight colours. This we
have checked by explicit calculation and again by making use of the fact that
$N_f^\mathrm{I}$ increases if any Dynkin index is increased.

As $N_f^\mathrm{I}>N_f^\mathrm{II}>N_f^\mathrm{III}$ the above findings are
handed down to the lower bound for the conformal window and the existence of
a Banks-Zaks fixed point.


\subsubsection{One flavour}

When the requirement is weakened to asymptotic freedom with one flavour,
all the above steps can be repeated after choosing new reference values:
%
$N_f^\mathrm{I}[\mathrm{A}_3({16})]=\frac{88}{91}$,
$N_f^\mathrm{I}[\mathrm{S}_3(5)]=\frac{55}{56}$,
$N_f^\mathrm{I}[\mathrm{R}_1(5)]=\frac{11}{14}$,
$N_f^\mathrm{I}[\mathrm{R}_2(7)]=\frac{77}{92}$,
$N_f^\mathrm{I}[\mathrm{R}_3(6)]=\frac{33}{38}$,
$N_f^\mathrm{I}[\mathrm{R}_4(5)]=\frac{11}{20}$.
%
This means that for the one flavour limit, no exceptions are present from 16
colours onward.


\subsection{Conformal window}

Evaluating the expression (\ref{ncrit}) for the lower bound of the conformal
window for all theories, which are asymptotically free with at least two
flavours, leads to the values listed in the fourth column of Table
\ref{nine}. For the fundamental, adjoint and two-index symmetric as well as
antisymmetric representations there exist theories with at least two
flavours, which have not yet entered the conformal phase. This remains true
beyond nine colours. From the exceptions below ten colours, the
three-index antisymmetric representation at nine colours and the four-index
antisymmetric representations are already conformal with less than two
flavours.

The minimal number of flavours necessary for a Banks--Zaks fixed point to
appear as calculated from Eq.~(\ref{bzfp}) is listed in the fifth column of
Table \ref{nine}. For values beyond nine colours, the explicit expressions
in Table \ref{table:reps} can be evaluated. The present analysis exhausts
the phase diagram for gauge theories with Dirac fermions in arbitrary
representations as function of the number of colours and flavours.

\begin{figure}[h]
\resizebox{8.5cm}{!}{\includegraphics{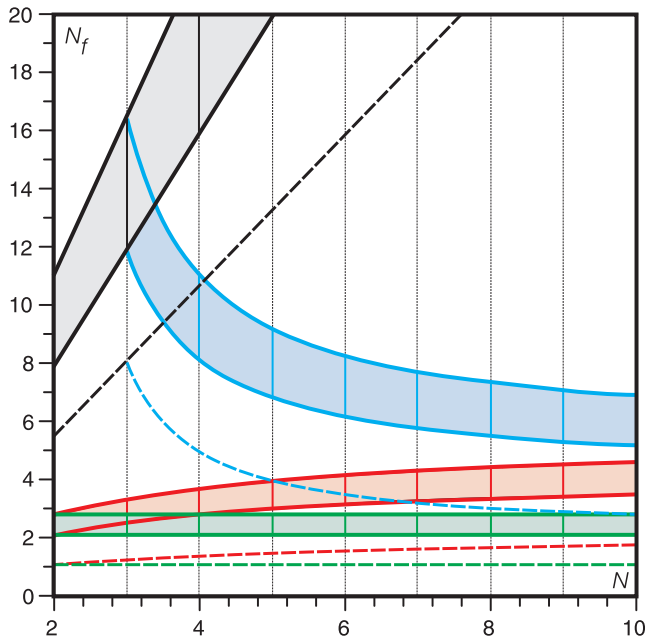}}
\caption{Phase diagram for theories with fermions in the (from top to bottom
in the plot; colour online): i) fundamental representation (grey), ii)
two-index antisymmetric (blue), iii) two-index symmetric (red), iv) adjoint
representation (green) as a function of the number of flavours and
the number of colours. The shaded areas depict the corresponding conformal
windows. The upper solid curve represents $N_f^\mathrm{I}[R(N)]$ (loss of
asymptotic freedom), the lower $N_f^\mathrm{II}[R(N)]$ (loss of chiral
symmetry breaking). The dashed curves show $N_f^\mathrm{III}[R(N)]$
(existence of a Banks--Zaks fixed point). Note how consistently the various
representations merge into each other when, for a specific value of $N$,
they are actually the same representation.}
\end{figure}


\section{Walking technicolour}

Here we will use our findings concerning asymptotically free theories with
at least one doublet of fermions to identify all physically acceptable 
technicolour models. In technicolour models the dynamical breaking of the 
electroweak symmetry from SU$_L$(2)$\times$U$_Y$(1) to U$_\mathrm{em}$(1) is 
not generated by an elementary Higgs particle like in the standard model, but 
by chiral symmetry breaking in an additional strongly interacting sector. It 
is made up of techniquarks transforming under the electroweak and an additional 
technicolour gauge group. The scale at which the chiral symmetry of the 
technicolour sector breaks is chosen to be the electroweak scale. Three of
the emergent Goldstone bosons are absorbed as longitudinal degrees of
freedom of the electroweak gauge bosons, which thus become massive. The
fermion masses are generated by embedding the electroweak and technicolour 
gauge groups in a larger extended technicolour gauge group. The gauge bosons 
of the extended technicolour model couple the fermions of the standard model 
to the techniquarks and their condensate, which renders the standard model
fermions massive.

Like all other mechanisms for electroweak symmetry breaking, technicolour has 
to face constraints derived from experimental data. In the case of 
technicolour the two main aspects are additional contributions to the
vacuum polarisation of the electroweak gauge bosons (oblique parameters) and 
flavour changing neutral currents as well as lepton number violation due to 
the extended technicolour dynamics. These issues have been discussed in great 
detail in the literature (see, for example \cite{Lane:2002wv,Hill:2002ap}). 
Experimental data (see, for example \cite{Eidelman:2004wy,unknown:2005em}) 
tells us that the above mentioned contributions must be small. Here, let us 
only recall that flavour changing neutral currents and lepton number violation 
are suppressed in walking technicolour theories, that is technicolour theories 
with nearly conformal dynamics. Through non-perturbative effects, 
quasi-conformality also helps reducing the techniquarks' contribution to the 
oblique parameters 
\cite{Sundrum:1991rf,Appelquist:1998xf,Appelquist:1999dq,Duan:2000dy,Hong:2006si}. 
(In the absence of quasi-conformal dynamics the $S$ parameter can be larger
than its perturbative value.)
On top of that, potential additional Goldstone bosons, beyond the three which 
are absorbed as the longitudinal degrees of freedom of the electroweak gauge 
bosons, become very heavy, thereby alleviating bounds set by them not having
been detected to date. Therefore, candidates for realistic technicolour 
theories should feature quasi-conformal dynamics and should contribute little 
to the oblique parameters already at the perturbative level. In what follows, 
we will quantify these criteria.

Already taking into account the experimental limits on the $S$-parameter
\cite{Peskin:1990zt} severely constrains the set of candidates.
Perturbatively, it is given by
\be
S=\frac{1}{6\pi}\frac{N_f}{2}d(\mathrm{R}).
\label{S}
\ee
The values for $S$ are given in Table \ref{nine}. Drawing the line at 
$S<\pi^{-1}$---somewhat arbitrarily but in accordance with the experimental 
limits \cite{Eidelman:2004wy,unknown:2005em}---leaves us with three 
candidates which, characterised by their Dynkin indices are: (1) with six 
flavours, (2) with two flavours, and (20) with two flavours. Doubling the 
value of the cut on the $S$ parameter ($S<2\pi^{-1}$) would admit two more: 
(11) with two flavours and (200) with two flavours.

The estimate for the lower bound (critical number of flavours) of the 
conformal window is based on the point where the critical coupling and the 
fixed point value coincide. This critical number of flavours is, in general,
not an even integer. A quasi-conformal physical realisation of a technicolour 
theory is, however, constructed from complete families of 
techniquarks.\footnote{Generalisations with an odd number of Dirac or even 
Weyl flavours are conceivable. A corresponding example is given in Sect. IIIC.}

\begin{figure*}[ht]
\qquad
\subfigure[~Technicolour, fully gauged]{
\resizebox{7cm}{!}{\includegraphics{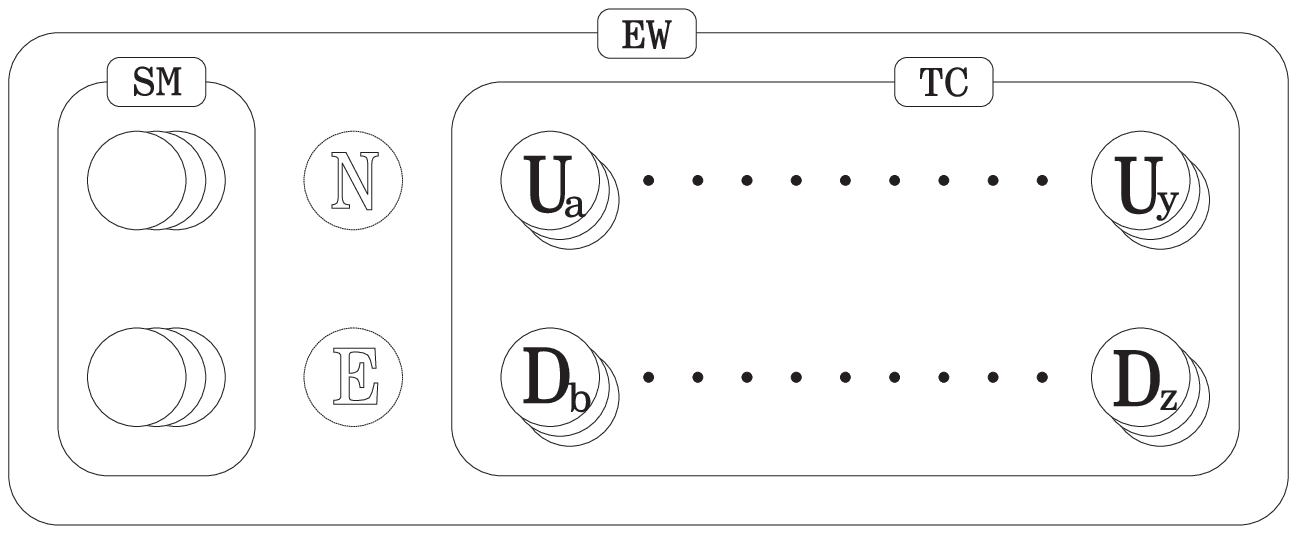}}
}
\qquad
\subfigure[~Technicolour, partially gauged]{
\resizebox{7cm}{!}{\includegraphics{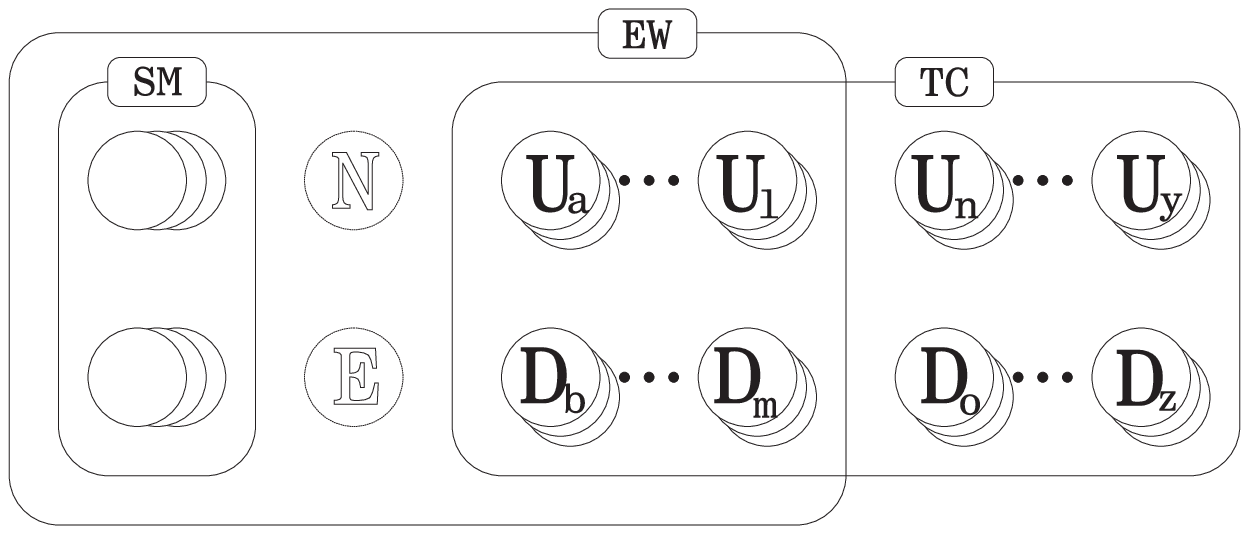}}
}
\qquad
\caption{\underline{Technicolour models.} The boxes depict under which gauge
groups the different particles transform. The box headlined "SM" represents
all standard model particles (excluding an elementary Higgs). N and E stand
for a fourth family of leptons, which may or may not have to be included in
order to evade a topological Witten anomaly. They have to be included, if
the number of techniquark families transforming under the electroweak symmetry
times the number of colour degrees of freedom is an odd number.
{\it Left panel:} In fully gauged
technicolour models, all techniquarks transform under the electroweak symmetry.
{\it Right panel:} In partially gauged technicolour a part of the techniquarks are
electroweak singlets. It is conceivable that only one (maximal splitting) or
several (sub-maximal splitting) families of techniquarks carry electroweak
charges. The latter set-up may be an alternative cure from the Witten
anomaly not involving additional leptons.}
\end{figure*}

From the difference of the two scales, the amount of walking, that is the 
ratio of the scale can be estimated \cite{Hong:2004td,Appelquist:1998xf},
\be
\lambda_*\approx\exp(\pi/\sqrt{\alpha_*/\alpha_c-1}).
\label{walking}
\ee
$\lambda_*$ is the ratio of the scale from which onwards the coupling 
constant stays approximately constant divided by the scale for which it 
starts running again. For this walking mechanism to be effective it must 
typically cover several decades. Setting the cut at $\lambda_*>10^3$ leaves 
(2) with two flavours (see Table \ref{nine}). [If the weaker bound on the $S$ 
parameter is chosen also (11) with two flavours survives.] Weakening the 
requirement on the range of the walking to $\lambda_*>10^2$ leads to no 
supplementary candidates.


\subsubsection{Two flavours, SU(2), adjoint representation: (2)}

The technicolour theory with two techniquarks in the two-index 
symmetric/adjoint representation of SU(2), that is (2), has been studied in 
\cite{Dietrich:2005jn,Dietrich:2005wk,Gudnason:2006yj} 
and found to be in good agreement with experimental constraints. It has to 
contain an additional family of leptons in order to avoid the topological 
Witten anomaly\footnote{It does not allow for an odd number of families of 
fermions to transform under an SU(2) gauge group [here the SU$_L$(2) of the 
electroweak gauge group]. The one family of (2), however, corresponds to 
three \mbox{[dim(2)=3]} technicolour copies added to the previously even 
number of SM fermion 
families.}\cite{Witten:1982fp}. This model has a rich phenomenology owing to 
the SU(2$N_f$=4) flavour symmetry in the unbroken phase, which is enhanced 
because the matter transforms under the adjoint representation, which is real.
The breaking to SO(4) leads to nine Goldstone bosons, three of which become 
the longitudinal 
degrees of freedom of the electroweak gauge bosons. The remaining six are of 
technibaryonic nature and may be very massive due to the intense walking of 
the theory. They can contribute to dark matter as might one of the additional 
leptons. This always depends on the hypercharge assignment for the particles 
beyond the standard model, which here is not fixed totally by requiring the 
freedom from gauge anomalies. Since, in the present theory, the fermions
transform under the adjoint representation, their technicolour can be
neutralised directly by technigluons, which leads to potentially 
phenomenologically interesting states. They are expected to have a mass of
the order of the confining scale of the theory.
For more details on this particular candidate see
\cite{Gudnason:2006yj}.


\subsubsection{Two flavours, SU(3), adjoint representation: (11)}

The theory with two techniflavours in the adjoint representation of SU(3),
(11), features a similarly enhanced flavour symmetry, SU(2$N_f$=4), whose
breaking to SO(4) gives rise to a total of nine Goldstone bosons with the same
consequences as in the two-colour case. The theory does not have to contain 
additional leptons whence the hypercharge assignment is fixed such that the 
techniquarks carry half-integer electrical charges with opposite signs. All 
objects of
two and more techniquarks can be technicolour neutral and the technicolour
of even a single techniquark can be neutralised by technigluons. This is in
direct analogy to the two-technicolour case.


\subsection{Partially gauged technicolour\label{pgt}}

A small modification of the traditional technicolour approach, which neither
involves additional particle species nor more complicated gauge groups, 
allows constructing several other viable candidates. It consists in letting 
only one doublet of techniquarks transform non-trivially under the electroweak
symmetries with the rest being electroweak singlets, as already suggested in 
\cite{Dietrich:2005jn} and later used in \cite{Christensen:2005cb}.
Still, all techniquarks transform under the technicolour gauge group. Thereby, 
perturbatively, only one techniquark doublet contributes to the oblique 
parameter which is thus kept to a minimum for theories which need more
than one family of techniquarks to be quasi-conformal. It is the condensation 
of that first electroweakly charged family that breaks the electroweak 
symmetry. Additionally, the number of ungauged techniquarks, in general, need 
not be even. In certain cases, this allows to come closer to conformality
(see Sect III B 1).
There exist several more partially gauged candidates with $S<\pi^{-1}$ 
(see Table \ref{nine}): (1) with seven flavours [instead of six fully gauged
flavours], (10) with eleven flavours, (100) with 15 flavours, (010) 
with eight flavours, (1000) with 19 flavours, and (10000) with 23 flavours.
Of these last-mentioned theories with techniquarks in the
fundamental representation, (10\dots0), only those with four or more colours
walk over more than two decades. Three decades could be reached from seven 
technicolours and 27 techniflavours onwards, which leads to a perturbative
contribution to the $S$ parameter of $\pi S\approx 1.2$ from the two
electroweakly gauged technifermions.
Therefore, the parametrically admissible models in the fundamental
representation are rather non-minimalist.

The theory with eight techniquarks in the two-index
antisymmetric representation of SU(4), (010), on the other hand, according to
Eq.~(\ref{walking}) walks over more than five decades. 

The techniquarks which are uncharged under the electroweak gauge group are
natural building blocks for components of dark matter.


\subsubsection*{Eight flavours, SU(4), two-index antisymmetric
representation: (010)}

Among the partially gauged cases the prime candidate is the theory with eight techniflavours
in the two-index antisymmetric representation of SU(4). The techniquarks of 
one of the four families carry electroweak charges while the others are 
electroweak singlets. Gauge anomalies are avoided if the two electrically 
charged techniquarks possess half-integer charges.
The technihadron spectrum can contain technibaryons made of only two
techniquarks because in the two-index
antisymmetric representation of SU(4) a singlet can already be formed in
that case [$(010)\otimes(010)\rightarrow(000)\oplus(101)\oplus(020)$].
Otherwise technisinglets can also be formed from four techniquarks. All 
technihadrons formed from techniquarks without electrical charges
can contribute to dark matter.
Due to the special charge assignment of the electrically charged particles
(opposite half-integer charges) certain combinations of those can also be 
contained in electrically uncharged technibaryons. For instance we can 
construct the following completely neutral technibaryon:
\be
\epsilon_{t_1t_2t_3t_4}Q_L^{{t_1t_2},f}Q_L^{{t_3t_4},f^{\prime}}
{\epsilon_{ff^{\prime}}} \ ,
\ee
where $\epsilon_{ff^{\prime}}$ saturates the SU(2)$_L$ indices of the two
gauged techniquarks and the first antisymmetric tensor $\epsilon$ is summed 
over the technicolour indices. We have suppressed the spin indices. If there 
is no technibaryon asymmetry the cross section for annihilation would be too
big and the present relic density would be negligible. In the presence of a 
technibaryon asymmetry, however, this particle would be another candidate for 
dark matter (different from those formed from electroweakly neutral
techniquarks), hardly detectable in any earth based experiment
\cite{Gudnason:2006yj}.

As (010) is a real representation the model's flavour symmetry is enhanced
to SU(2$N_f$=16)\footnote{It is slightly explicitly broken by the
electroweak interactions. This is, of course, always the case between the
techniups and technidowns, but here, additionally, there is a difference, on
the electroweak level, between gauged and ungauged techniquark families.}, 
which, when it breaks to SO(16), induces 135 Goldstone
bosons\footnote{Obviously the Goldstone must receive a sufficiently large
mass. This is usually achieved in extended technicolour. Still, they could be
copious at LHC.}. 

It is worth recalling that the centre group symmetry left invariant
by the fermionic matter is a Z$_2$
symmetry. Hence there is a well defined order parameter for confinement
\cite{Sannino:2005sk} which can play a
role in the early Universe.


\subsection{Beyond the prime candidate}

In view of the uncertainty of the experimental limits for the oblique
parameters (see, for example \cite{Dutta:2006cd}) and the approximations made
here to determine the degree of walking, not all the other theories, which
above have not been identified as prime candidates for a realistic walking
technicolour theory, should be considered as being ruled out. While the
theories which, in this analysis, stood out as most favoured, would, with the
highest likelihood, survive an amelioration of the experimental limits
and/or a refinement of the theoretical assessment, others might emerge as
being viable candidates as well. For these reasons, we address some of those
in the rest of this section. At the end, we will have addressed all settings
listed in Table \ref{tc}.

\begin{table}[h]
\begin{tabular}{c|c|c|c|c|c|c}
R&$N_f$&flavour symmetry&WA&P-G&$\lg\lambda_*$
&$\pi S$\footnote{The contribution to $S$ from additional leptons is not 
included. It is of the order $\pi S\approx 1/6$. The contribution to 
the $T$ parameter from non-mass-degenerate leptons helps adjusting 
to the correlation of the $S$ and $T$ parameters indicated by experimental
data \cite{Dietrich:2005jn,Dietrich:2005wk}.}\\
\hline (1)&7&SU(14)$\rightarrow$Sp(14)&no&yes&1.3
&0.3\footnote{This value results for a partial weak gauging
involving only one doublet of the theory. If all possible, that is
three doublets are gauged one has $\pi S\approx1$.}\\
\bf{(2)}&2&SU(4)$\rightarrow$SO(4)&yes&--&3.1&0.5\\
(10)&11&SU$_L$(11)$\times$SU$_R$(11)
&yes&yes\footnote{In
these cases, the global Witten anomaly can be
circumvented either by adding one family of leptons or by having two families of techniquarks charged under the
electroweak interactions. The value for the $S$ parameter is for the former
case. In the latter, it is twice as large.}
&1.8&0.5\\
(20)&2&SU$_L$(2)$\times$SU$_R$(2)
&no&--&1.1&1.0\\
(11)&2&SU(4)$\rightarrow$SO(4)&no&--&3.1&1.3\\
(100)&15&SU$_L$(15)$\times$SU$_R$(15)
&no&yes&2.2&0.7\\
(200)&2&SU$_L$(2)$\times$SU$_R$(2)&no&--&0.8&1.7\\
\bf{(010)}&8&SU(16)$\rightarrow$SO(16)
&no&yes&5.1&1.0\\
(101)&2&SU(4)$\rightarrow$SO(4)&yes&--&3.1&2.5\\
(1000)&19&SU$_L$(19)$\times$SU$_R$(19)&yes&yes$^c$&2.4&0.8\\
(0100)&6&SU$_L$(6)$\times$SU$_R$(6)
&no&yes&1.5&1.7\\
(10000)&23&SU$_L$(23)$\times$SU$_R$(23)&no&yes&2.7&1.0\\
\end{tabular}
\caption{List of possible technicolour theories. The first two columns give 
the representation and the number of flavours, respectively.
In the next column follows the symmetry breaking pattern.
[SU$_L$($N_f$)$\times$SU$_R$($N_f$) always breaks to SU($N_f$).]
Thereafter, it is marked whether a global Witten anomaly has to be taken care 
of by adding a family of leptons or, where this is possible, by a less 
minimal partial weak gauging of the theory. Finally, the amount of walking 
and the perturbative $S$ parameter are listed. The prime candidates, that is 
the theories with the largest amount of walking at an acceptable size for the 
oblique corrections are marked in boldface.} \label{tc}
\end{table}

\subsubsection{Fundamental}

Compared to the prime examples with matter in higher-dimensional
representations, models with fundamental matter must have a comparatively
large number of colours and flavours in order to feature a sufficient amount
of walking. In the partially gauged approach their $S$ parameter reaches
$\pi^{-1}$ for six colours. Phenomenologically, the models with fundamental
techniquarks can be divided into two groups with an even, respectively odd
number of colours (=dimension). Technicolour singlets are always formed by 
$N$ techniquarks.

For an even number of colours the model with one electroweakly gauged
techniquark doublet is free of the Witten anomaly. No additional lepton
family has to be included. Opposite half-integer electrical charges for the
technifermions avoid gauge anomalies. Technicolour singlets constructed from
the techniquarks which are gauged under the electroweak interactions carry
integer electrical charges. This also includes uncharged technibaryons which 
can be components of dark matter.

For an odd number of colours a single doublet of techniquarks gauged under
the electroweak interactions leads to a Witten anomaly. It can be
circumvented by including one additional lepton family. This allows for a
more general hypercharge assignment \cite{Dietrich:2005jn} than in the case
without leptons.

An alternative to an additional lepton family for circumventing a Witten 
anomaly is to gauge two instead of one techniquark doublet, at the cost of
doubling the $S$ parameter. For this non-minimal weak gauging the
hypercharge assignment which corresponds to opposite half-integer electric 
charges for the techniquarks avoids gauge anomalies.  


\subsubsection{Two-index symmetric}

Two more theories with apparently insufficient walking but which make at
least the weaker 
bounds on the $S$ parameter are the two flavour theories with techniquarks in 
the two-index symmetric representation of SU(3), (20), and SU(4), (200). None 
of the two needs additional leptons. Thus the hypercharge assignment is 
completely fixed. Three respectively four techniquarks are needed to form a 
singlet technibaryon.


\subsubsection{Two flavours, SU(4), adjoint}

The four technicolour model with two techniflavours in the adjoint
representation has a relatively large perturbative $S$ parameter but walks
over more than three decades which helps reducing it through
non-perturbative corrections. It also requires an additional lepton family
in order to cure the Witten anomaly, which allows for a more general
hypercharge assignment. The enhanced symmetry, SU(2$N_f$=4), breaks to SO(4)
and leaves behind nine Goldstone bosons. They must be rendered massive which
also necessitates a good amount of walking. As for all other adjoint models,
any number of techniquarks can form a singlet and technigluons
can neutralise technicolour as well.


\subsubsection{Six flavours, SU(5), two-index antisymmetric}

Finally, let us mention the five technicolour model with six techniflavours 
in the two-index antisymmetric representation. For an acceptable $S$ 
parameter, only part of the techniflavours can be gauged under the
electroweak interactions. No additional leptons are required.
This variant does not exhibit a remarkable amount of walking.


\subsection{Split technicolour: A review}

Since one aim of this work is to provide a catalogue of various possible 
walking type, non-supersymmetric gauge theories, which can be used to 
dynamically break the electroweak symmetry we summarize here also another
possibility already appearing in \cite{Dietrich:2005jn}. There we also 
suggested a way of keeping the technifermions in the fundamental 
representation while still reducing the number of techniflavours needed to be
near the conformal window. Like for the partially gauged case described above 
this can be achieved by adding matter uncharged under the weak interactions. 
The difference to section \ref{pgt} is that this part of matter transforms 
under a different representation of the technicolour gauge group than the 
part coupled directly to the electroweak sector. We choose it to be a 
massless Weyl fermion in the adjoint representation of the technicolour gauge 
group, that is a technigluino. 
The resulting theory has the same matter content as $N_f$-flavour super QCD but without the scalars;
hence the name "split technicolour." We expect the critical number of flavours above which one enters the
conformal window $N_f^\mathrm{II}$ to lie within the range
\be
\frac{3}{2}<\frac{N^\mathrm{II}_f}{N}<\frac{11}{2} \ .
\ee
The lower bound is the exact supersymmetric value for a non-perturbative
conformal fixed point \cite{Intriligator:1995au}, while the upper bound is
the one expected in the theory without a technigluino. The matter content of
"split technicolour" lies between that of super QCD and the standard
fundamental technicolour theory.

For two colours the number of (techni)flavours needed to be near the
conformal window in the split case is at least three, while for three
colours more than five flavours are required. These values are still larger
than the ones for theories with fermions in the two-index
symmetric representation. It is useful to remind the reader that in
supersymmetric theories the critical number of flavours needed to enter the
conformal window does not coincide with the critical number of flavours
required to restore chiral symmetry. The scalars in supersymmetric theories
play an important role from this point of view. We note that a split
technicolour-like theory has been used recently in \cite{Hsu:2004mf}, to
investigate the strong CP problem.

Split technicolour shares some features with theories of split
supersymmetry recently advocated and studied in
\cite{Arkani-Hamed:2004fb,{Giudice:2004tc}} as possible extensions of the
standard model. Clearly, we have introduced split technicolour---differently
from split supersymmetry---to address the hierarchy problem. This is why we
do not expect new scalars to appear at energy scales higher than the one of
the electroweak theory.


\section{Near Conformal Spectrum: Refinements.}

An interesting question to ask is how the spectrum of light particles looks
near the infrared fixed point. When these theories are used to dynamically 
break the electroweak symmetry this corresponds to asking, whether the
composite Higgs for walking theories is light or heavy with respect to the
electroweak symmetry breaking scale?

There is consensus that at least part of the hadronic spectrum becomes light near a conformal fixed point
\cite{Chivukula:1996kg}. This might not be too surprising after all, since one expects the chiral symmetry
breaking scale to vanish at the conformal fixed point. A more delicate issue
is, if part of the spectrum and
more specifically the chiral partner of the pion becomes light faster than the chiral symmetry breaking scale
as we tune the number of flavours toward the fixed point value \footnote{We thank T. Appelquist and K. Lane for
pressing on this relevant point.}.

We will use a simple analysis to elucidate why we expect to find a light composite scalar object,
---light with respect to the chiral symmetry breaking scale---near the conformal fixed point. If, as function of the
number of flavours, there is a smooth phase transition one applies Wilson's
approach and investigates the corresponding renormalisable scalar effective 
Lagrangian. In our analysis we are assuming
that the rest of the composite
spectrum of the theory decouples near this fixed point \footnote{Note that
Chivukula \cite{Chivukula:1996kg}
argued that the rest of the dynamics does not decouple near the same fixed
point.}. This is a standard
assumption if a smooth transition occurs. Here the situation is different
from the case of a thermal
phase transition since fermions do not necessarily decouple at the 
four-dimensional fixed point.

For a generic number of space-time dimensions $d$ one finds that the fixed point value of the $\phi^4$ coupling constant
$g/4!$ of the scalar effective theory with an $O(N_f)$ global symmetry 
is\footnote{This part is meant to be illustrative. If the reader is 
interested in more details on the fixed point analysis for theories with 
unitary flavour groups these can be found in \cite{Basile:2005hw}.},
\be 
g_{\ast} = \frac{48\pi^2}{N_f + 8} (4-d) \ . 
\ee
Hence, in four dimensions, the effective fixed point coupling constant vanishes. This is consistent with the
intuitive idea that near a conformal fixed point 
the presence of a non-zero
coupling would lead to the generation of a mass term driving the theory away from conformality. Using the
renormalisation group approach we compare the coupling constant at the chiral symmetry breaking scale 
$\Sigma_0$---that is $g(\Sigma_0)=\bar g$---with the same coupling at a much larger scale $\Lambda$ which we identify with the
scale above which the underlying coupling constant starts running again. We call the latter the bare coupling
$g$ and we have:
\be \bar{g} &\sim& \frac{g} { 1 + \frac{N_f + 8}{48\pi^2}g\left(\frac{\alpha_{\ast}}{\alpha_c}-1\right)^{-1/2} }
\sim \nn &\sim& \frac{48\pi^2}{N_f + 8} \left(\frac{\alpha_{\ast}}{\alpha_c}-1\right)^{1/2} \ .
\label{coupling}\ee
In the previous equation we used the following relation between the dynamically generated quark mass at zero
momentum, $\Sigma_0$, and the scale $\Lambda$ above which the underlying theory
starts running again:
\be \ln\left(\frac{\Sigma_0}{\Lambda}\right) \sim \left(\frac{\alpha_{\ast}}{\alpha_c}-1 \right)^{-1/2} \ . \ee
The function $(\alpha_{\ast}/\alpha_c -1)$ embodies the underlying dynamics,
and, hence, is sensitive to the
number of flavours, colours and the fermion representation with respect to the gauge 
group\footnote{Nota Bene: The coefficient in front of 
$(\alpha_{\ast}/\alpha_c -1)$ depends on the underlying fermion
representation with respect to the gauge group which dictates, ultimately, the global symmetry of the problem.}.
The physical mass of the scalar field in units of the vacuum expectation value of the scalar field is given by
\be \frac{M^2_{\phi}}{v^2} = 2\bar{g} \sim \left(\frac{\alpha_{\ast}}{\alpha_c} -1 \right)^{1/2}, \ee
with $v = \langle \phi \rangle\sim \Sigma_0$. Hence it vanishes near the 
transition faster than the vacuum expectation value, that is the chiral 
symmetry breaking scale. This supports the proposal that the composite
Higgs can be lighter for walking technicolour theories than for ordinary
technicolour theories away from
conformality\footnote{In our previous analysis \cite{Dietrich:2005jn} we 
suggested a dependence of the scalar mass on the number of 
flavours near the fixed point, using trace anomaly arguments. This
result, however, did not take into account the scale dependence of the 
$\phi^4$ coupling.}.

There are some caveats related to the analysis above. If the transition is not 
smooth the present analysis is simply not applicable. This is the conclusion
drawn, for example in \cite{Appelquist:1996dq} and in 
\cite{Gies:2002hq,Gies:2005as}. Note that this uncertainty is present in every 
study of a phase transition via effective theories. Another issue is that for 
zero temperature phase transitions fermions are no longer screened near the 
fixed point and should be included in the analysis above. Besides, we could 
have also terms related to the underlying axial anomaly, depending on the 
number of flavours and the global flavour symmetry. However, if the phase 
transition happens near two flavours one expects the axial anomaly, 
which is related to a marginal operator, not to dominate the fixed point 
dynamics. Finally, we have also assumed valid the low energy effective 
description all the way to the scale $\Lambda$.


\section{Summary}

We have presented a comprehensive analysis of the phase diagram of
non-supersymmetric vector-like and strongly coupled SU($N$) gauge theories 
with matter in various representations of the gauge group. We have considered 
models with fermions in a single representation of the gauge group, but also
the case of a combination of fermions in the fundamental with fermions in the 
adjoint representation. As physical application we considered the dynamical 
breaking of the electroweak symmetry via walking technicolour.
We have then taken into account constraints form electroweak precision 
measurements and, thereby, reduced the number of theories viable for 
correctly describing the dynamical breaking. Still, we
find that a considerable number of strongly coupled theories are 
still excellent candidates for breaking the electroweak symmetry dynamically.


\section*{Acknowledgements}

It is a pleasure to thank J.~Schechter, K.~Splittorff, and 
K.~Tuominen for discussions. The work of F.S. is supported by the Marie Curie
Excellence Grant under contract MEXT-CT-2004-013510 as well as the Danish Research
Agency.


\appendix


\section{Basic group theory relations \label{eleven}}

The Dynkin indices label the highest weight
of an irreducible representation and uniquely characterise the
representations. The Dynkin indices for some of the most common
representations are given in Table \ref{dynkin}. For details on the concept
of Dynkin indices see, for example \cite{Dynkin:1957um,Slansky:1981yr}.

\begin{table}
\begin{tabular}{r|c}
representation&Dynkin indices\\
\hline
singlet&(000\dots 00)\\
fundamental (F)&(100\dots 00)\\
antifundamental ($\bar{\mathrm{F}}$)&(000\dots 01)\\
adjoint (G)&(100\dots 01)\\
$n$-index symmetric (S$_n$)&(n00\dots 00)\\
2-index antisymmetric (A$_2$)&(010\dots 00)\\
\end{tabular}
\caption{Examples for Dynkin indices for some common representations.}
\label{dynkin}
\end{table}

For a representation, R, with the Dynkin indices
$(a_1,a_2,\dots,a_{N-2},a_{N-1})$ the
quadratic Casimir operator reads \cite{White:1992aa}
\be C_2(\mathrm{R}) &=& \sum_{m=1}^{N-1}[ N(N-m)ma_m + m(N-m){a_m}^2 +\nonumber \\&&
\sum_{n=0}^{m-1}2n(N-m)a_na_m ] \label{C} \ee
and the dimension of R is given by
\be
d(\mathrm{R})
=
\prod_{p=1}^{N-1}
\left\{
\frac{1}{p!}
\prod_{q=p}^{N-1}
\left[
\sum_{r=q-p+1}^p(1+a_r)
\right]
\right\},
\label{d}
\ee
which gives rise to the following structure
\be d(\mathrm{R}) &=& (1+a_1)(1+a_2)\dots(1+a_{N-1}) \times
\nn && \times (1+\sfr{a_1+a_2}{2}) \dots
(1+\sfr{a_{N-2}+a_{N-1}}{2}) \times
\nn && \times (1+\sfr{a_1+a_2+a_3}{3}) \dots
(1+\sfr{a_{N-3}+a_{N-2}+a_{N-1}}{3}) \times
\nn && \times \dots \times
\nn && \times
(1+\sfr{a_1+\dots+a_{N-1}}{N-1}). \ee
The Young tableau associated to a given Dynkin index
$(a_1,a_2,\dots,a_{N-2},a_{N-1})$ is easily constructed. The length of
row $i$ (that is the number of boxes per row) is given in terms of the Dynkin
indices by the expression $r_i = \sum_{i}^{N-1} a_{i}$. The length of
each column is indicated by $c_k$; $k$ can assume any positive integer value.
Indicating the total number of boxes associated to a given Young tableau with 
$b$ one has another compact expression for $C_2(\mathrm{R})$,
\be
C_2(\mathrm{R})
= 
N\left[ bN +\sum_i r^2_i  -\sum_ic_i^2 - \frac{b^2}{N} \right]\ ,
\ee
and the sums run over each column and row.



\end{document}